\documentclass[conference, romanappendices,10pt]{IEEEtran}
\IEEEoverridecommandlockouts

\usepackage{amsmath,amssymb,amsthm,bbm,cite,color,enumitem,flushend,graphicx,microtype,setspace,url}
\usepackage{mleftright}
\usepackage{graphicx}
\usepackage{tikz}
\usepackage[USenglish]{babel}
\usepackage[utf8]{inputenc}
\usepackage[T1]{fontenc}
\usepackage[algo2e]{algorithm2e}
\usepackage{algorithm}
\usepackage{comment}
\usepackage{epstopdf}

\renewcommand{\d}{\mathbf{d}}

\newcommand{\h}{\mathbf{h}}

\newcommand{\p}{\mathbf{p}}

\newcommand{\w}{\mathbf{w}}

\newcommand{\D}{\mathbf{D}}

\renewcommand{\H}{\mathbf{H}}
\newcommand{\I}{\mathbf{I}}

\renewcommand{\P}{\mathbf{P}}
\newcommand{\Q}{\mathbf{Q}}
\newcommand{\R}{\mathbf{R}}
\renewcommand{\S}{\mathbf{S}}

\newcommand{\W}{\mathbf{W}}

\newcommand{\Y}{\mathbf{Y}}
\newcommand{\Z}{\mathbf{Z}}

\newcommand{\Psib}{\mathbf{\Psi}}
\newcommand{\Omegab}{\mathbf{\Omega}}

\newcommand{\setB}{\mathcal{B}}
\newcommand{\setC}{\mathcal{C}}
\newcommand{\setD}{\mathcal{D}}

\newcommand{\setK}{\mathcal{K}}

\newcommand{\setN}{\mathcal{N}}

\newcommand{\Real}{\mbox{$\mathbb{R}$}}
\newcommand{\Compl}{\mbox{$\mathbb{C}$}}

\newcommand{\argmin}{\operatornamewithlimits{argmin}}

\newcommand{\Exp}{\mathbb{E}}
\newcommand{\herm}{\mathrm{H}}

\renewcommand{\Re}{\mathrm{Re}}

\newcommand{\tr}{\mathrm{tr}}
\newcommand{\tran}{\mathrm{T}}

\newcommand{\fro}{\mathrm{F}}

{}
{}
{}
{}
{}
{}
{}

\newcommand{\pilot}{\textnormal{\tiny{p}}}
\newcommand{\data}{\textnormal{\tiny{d}}}

\usepackage{tikz}
\usetikzlibrary{arrows,backgrounds,calc,intersections,decorations.pathreplacing,positioning,shapes}
\usepackage{pgfplots}
\pgfplotsset{compat=1.18} 
\title{Data-Aided Regularization of Direct-Estimate Combiner in Distributed MIMO Systems}\author{
\IEEEauthorblockN{Bikshapathi Gouda, Italo Atzeni, and Antti Tölli}
\IEEEauthorblockA{Centre for Wireless Communications, University of Oulu, Finland \\
E-mail: \{bikshapathi.gouda, italo.atzeni, antti.tolli\}@oulu.fi 
\thanks{This work was supported by the Academy of Finland (336449 Profi6, 348396 HIGH-6G, 357504 EETCAMD, and 369116 6G~Flagship) and by the European Commission (101095759 Hexa-X-II).}}}

\begin{document}


\maketitle

\begin{abstract}
This paper explores the data-aided regularization of the direct-estimate combiner in the uplink of a distributed multiple-input multiple-output system. The network-wide combiner can be computed directly from the pilot signal received at each access point, eliminating the need for explicit channel estimation. However, the sample covariance matrix of the received pilot signal that is used in its computation may significantly deviate from the actual covariance matrix when the number of pilot symbols is limited. To address this, we apply a regularization to the sample covariance matrix using a shrinkage coefficient based on the received data signal. Initially, the shrinkage coefficient is determined by minimizing the difference between the sample covariance matrices obtained from the received pilot and data signals. Given the limitations of this approach in interference-limited scenarios, the shrinkage coefficient is iteratively optimized using the sample mean squared error of the hard-decision symbols, which is more closely related to the actual system's performance, e.g., the symbol error rate (SER). Numerical results demonstrate that the proposed regularization of the direct-estimate combiner significantly enhances the SER, particularly when the number of pilot symbols is limited.
\smallskip

\end{abstract}

\section{Introduction} \label{sec:INTRO}
Distributed multiple-input multiple-output (MIMO) systems consist of geographically separated access points (APs) connected to a central unit (CU) via fronthaul links for data and channel state information (CSI) exchange. The APs coherently serve the user equipments (UEs) by utilizing a network-wide precoder or combiner with the objective of improving the spectral efficiency and providing a uniform service across the network~\cite{Ngo17, Atz21, Gou24}. Focusing on the uplink, the network-wide combiner can be computed directly from the pilot signal received at each AP without the need for explicit channel estimation. This is referred to as \textit{direct-estimate combiner}, which inherently accounts for potential interference by utilizing the sample covariance matrix of the received pilot signal~\cite{Gou24, Atz21, Shi14}. In the absence of statistical CSI, the least-squares (LS) method can be used to compute the direct-estimate combiner. When the number of pilot symbols is limited, a regularization of the sample covariance matrix can be employed to enhance the system's performance. For instance, the use of exponentially weighted regularization across multiple estimates was investigated in~\cite{Shi14} to improve the accuracy of the direct-estimate precoder and combiner under time-varying channels.

Many studies have focused on accurately estimating the interference-plus-noise covariance matrix using received signal samples. Shrinkage techniques have been used to adjust the diagonal elements with a weighted factor determined by the shrinkage coefficient~\cite{Che11, Ran21}. For instance, the impact of different shrinkage coefficients is analyzed in~\cite{Dav20}. Additionally, methods like leave-one-out cross-validation can be employed to further enhance the accuracy of the sample covariance matrix~\cite{Ton16, Ton16b}. However, the estimation accuracy of the covariance matrix is usually not directly related to metrics such as the symbol error rate (SER) or signal-to-interference-plus-noise ratio (SINR). To address this, an expected likelihood criterion was proposed in~\cite{Kuz16} to optimize the shrinkage coefficient, demonstrating effectiveness across a range of SINR values with varying numbers of UEs and interference sources. Furthermore, the shrinkage coefficient was specifically adjusted to maximize the SINR in~\cite{Qia24}.

This paper considers a distributed MIMO system serving multiple UEs in the uplink. The network-wide combiner is computed at the CU via direct estimation by minimizing the LS objective over the received pilot signal. Due to the limited number of pilot symbols, the sample covariance matrix used in the computation of the direct-estimate combiner may be inaccurate, leading to a degradation of the system's performance. To address this, we apply a regularization to the sample covariance matrix by adjusting its diagonal elements with a shrinkage coefficient. Assuming that the interference statistics do not change during both the pilot and data transmission phases, the shrinkage coefficient is optimized using the received data signal as follows. Initially, a closed-form expression of the shrinkage coefficient is derived by minimizing the difference between the sample covariance matrices obtained from the received pilot and data signals. However, recognizing the limitations of this approach in interference-limited scenarios, we propose an iterative optimization of the shrinkage coefficient that minimizes the sample mean squared error (MSE) of the hard-decision symbols. In this approach, the shrinkage coefficient is iteratively updated using a gradient-based method along with the hard-decision symbols. Numerical results demonstrate that the regularization of the direct-estimate combiner significantly enhances the SER in interference-limited scenarios and with a small number of pilot symbols.

\section{System Model}
We consider a set of APs $\setB \triangleq \{1, \ldots, B\}$, each equipped with $M$ antennas, serving a set of single-antenna UEs $\setK \triangleq \{1, \ldots, K\}$ in the uplink. During the data transmission phase, the received signal at AP~$b$ is given by
\begin{align} \label{eq:data}
    \Y^{\data}_{b} \triangleq \sum_{k \in \setK} \sqrt {\rho_k } \h_{b,k} \d^{\herm}_k + \Z^{\data}_b \in \Compl^{M\times \tau^{\data}},
\end{align}
where $\rho_k \in \Real_{+}$ is the transmit power of UE~$k$, $\h_{b,k} \in \Compl^{M \times 1}$ is the uplink channel between UE~$k$ and AP~$b$, $\d_k \in \Compl^{ \tau^{\data} \times 1}$ is the data symbol vector transmitted by UE~$k$ (with $\Exp \big[ \|\d_k \|^{2} \big] = \tau^{\data}$), and $\Z^{\data}_b \in \Compl^{M \times \tau^{\data}}$ represents the combination of interference from out-of-cluster UEs (i.e., UEs served by other APs) and additive white Gaussian noise (AWGN) at AP~$b$. Accordingly, the aggregated received data signal  across all the APs is
\begin{align}
    \Y^{\data} & \triangleq \big[(\Y^{\data}_{1})^{\tran}, \ldots, (\Y^{\data}_{B})^{\tran}\big]^{\tran} \nonumber \\
    & = \sum_{k \in \setK} \sqrt {\rho_k } \h_{k} \d^{\herm}_k + \Z^{\data} \in \Compl^{BM\times \tau^{\data}}, \label{eq:agg_data}
\end{align}
with $\h_{k} \triangleq \big[\h_{1,k}^{\tran}, \ldots, \h_{B,k}^{\tran}\big]^{\tran} \in \Compl^{B M \times 1}$ and $\Z^{\data} \triangleq \big[(\Z^{\data}_{1})^{\tran}, \ldots, (\Z^{\data}_{B})^{\tran}\big]^{\tran} \in \Compl^{B M \times \tau^{\data}}$, where each column of $ \Z^{\data}$ is assumed to follow a circularly-symmetric complex normal distribution with zero mean and covariance matrix $\Psib \in \Compl^{BM\times BM}$. The aggregated received data signal in \eqref{eq:agg_data} is obtained at the CU by combining the received signals in \eqref{eq:data} forwarded by each AP via fronthaul links.

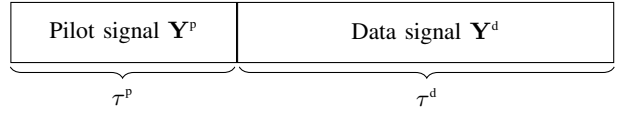
\begin{figure}
\begin{center}
\begin{tikzpicture}
\usetikzlibrary{calc}
\newdimen\dis
\dis=2cm;
\small
\tikzstyle{block} = [rectangle, draw, text centered, minimum height=0.4\dis]

\node [block, minimum width=1.5\dis] (slot0){{Pilot signal $\Y^{\pilot}$}};
\node [block, right=0\dis of slot0, minimum width=2.5\dis] (slot1) {{Data signal $\Y^{\data}$}};

\draw [decorate, decoration={brace, mirror, raise=0.5mm, amplitude=1.5mm}] ($(slot0.south west)+(0mm,0mm)$) -- ($(slot0.south east)+(-0.3mm,0mm)$) node [midway, xshift=0mm, yshift=-4.5mm, align=center] {$\tau^{\pilot}$};
\draw [decorate, decoration={brace, mirror, raise=0.5mm, amplitude=1.5mm}] ($(slot1.south west)+(0.3mm,0mm)$) -- ($(slot1.south east)+(0mm,0mm)$) node [midway, xshift=0mm, yshift=-4.5mm, align=center] {$\tau^{\data}$};
\end{tikzpicture}
\vspace{-5mm}
\end{center}
\caption{Uplink transmission of the pilot and data signals within the coherence interval.}
\label{fig:ul_plt_data}
\vspace{-6mm}
\end{figure}

The soft estimate of $\d_k$ can be obtained at the CU by combining $\Y^{\data}$ with the combiner $\w_k \in \Compl^{BM \times 1} $ as
\begin{align}\label{eq:d_k_est}
    \hat \d_k \triangleq (\Y^{\data})^{\herm} \w_k \in \Compl^{ \tau^{\data} \times 1}.
\end{align}
Since the computation of the combiner $\w_k$ requires CSI at the CU, we assume that each UE transmits pilots for the channel estimation along with the data in each coherence interval, as illustrated in Figure~\ref{fig:ul_plt_data}. Let $\p_k \in \Compl^{\tau^{\pilot} \times 1}$ represent the pilot vector transmitted by UE~$k$ (with $\|\p_k\|^2 = \tau^{\pilot}$). During the pilot transmission phase, the received signal at AP~$b$ is given~by
\begin{align}
 \Y^{\pilot}_{b} & \triangleq \sum_{k \in \setK}\sqrt{\rho_k } \h_{b,k}\p_k^\herm + \Z^{\pilot}_b \in \Compl^{M\times\tau^{\pilot}},
 \end{align}
where $\Z^{\pilot}_b$ represents the combination of interference from out-of-cluster UEs and AWGN at AP~$b$ (similar to $\Z^{\data}_b$ in the data transmission phase). Consequently, the aggregated received pilot signal across all the APs is
\begin{align}
    \Y^{\pilot} & \triangleq \big[(\Y^{\pilot}_{1})^{\tran}, \ldots, (\Y^{\pilot}_{B})^{\tran}\big]^{\tran} \nonumber \\
                & =   \sum_{k \in \setK} \sqrt {\rho_k } \h_{k} \p_k^{\herm} + \Z^{\pilot} \nonumber\\
                & = \H \Omegab_{\rho}^{\frac{1}{2}} \P^{\herm} + \Z^{\pilot} \in \Compl^{BM \times \tau^{\pilot}}, \label{eq:agg_plt}
\end{align}
with $\H \triangleq [\h_1, \ldots, \h_K] \in \Compl^{B M \times K}$, $\Omegab_{\rho} \triangleq\textrm{Diag}\big([{\rho_1},\ldots {\rho_K]}\big) \in \Real^{K \times K}$, $\P \triangleq [\p_1,\ldots, \p_K] \in \Compl^{\tau^{\pilot} \times K}$, and $\Z^{\pilot} \triangleq \big[(\Z^{\pilot}_{1})^{\tran}, \ldots, (\Z^{\pilot}_{B})^{\tran}\big]^{\tran} \in \Compl^{BM \times \tau^{\pilot}}$. Assuming that the interference characteristics remain constant during both the pilot and data transmission phases, each column of $\Z^{\pilot}$ follows the same distribution as the columns of $\Z^{\data}$. In the following, we discuss the computation of the direct-estimate combiner and its regularization using the received pilot and data signals.

\section{Regularization of the Direct-Estimate Combiner}
In the absence of statistical CSI, the aggregated direct-estimate combiner for all the UEs, denoted by $\W \triangleq [\w_1, \ldots, \w_K] \in \Compl^{B M \times K}$, can be obtained by minimizing the LS objective over the received pilot signal, i.e.,
\begin{align}
    \underset{\W}{\argmin} \, \|\W^{\herm} \Y^{\pilot} - \P^{\herm}\|_{\fro}^2 = \frac{1}{\tau^{\pilot}} \Q^{-1} \Y^{\pilot} \P,
\end{align}
with $\Y^{\pilot}$ in \eqref{eq:agg_plt} and where \( \Q \triangleq \frac{1}{\tau^{\pilot}}\Y^{\pilot} (\Y^{\pilot})^{\herm} \in \Compl^{B M \times B M} \) is the sample covariance matrix of the received pilot signal, which converges to the actual covariance matrix \( \H \Omegab_{\rho} \H^{\herm} + \Psib \) as \( \tau^{\pilot} \rightarrow \infty \). However, with a limited sample size (i.e., with small \( \tau^{\pilot} \)), \( \Q \) may not accurately approximate the actual covariance matrix~\cite{Che11}. In this setting, a regularization of the sample covariance matrix using shrinkage techniques can be applied to improve the estimation accuracy. Specifically, we adjust the diagonal elements of \( \Q \) with a shrinkage coefficient \( \alpha \in (0, 1] \) as~\cite{Che11}
\begin{align}
    \R(\alpha) &\triangleq (1-\alpha)\Q + \alpha \frac{\tr(\Q)}{BM} \I_{BM}, \nonumber \\
       & = \Q + \alpha \S \in \Compl^{B M \times B M}, \label{eq:cov_shrink}
\end{align}
with \( \S \triangleq \frac{\tr(\Q)}{BM} \I_{BM} - \Q  \in \Compl^{B M \times B M} \). In \eqref{eq:cov_shrink}, the regularization shrinks the sample covariance matrix towards a diagonal matrix to improve the estimation accuracy based on the received pilot signal.\footnote{The shrinkage coefficient can vary for each AP; however, for simplicity, we use a single coefficient across all the APs and optimize it at the CU. Alternatively, it is also possible to estimate the shrinkage coefficient at each AP in a distributed fashion using local CSI.} Finally, with the regularized sample covariance matrix, the direct-estimate combiner can be rewritten as
\begin{align}\label{eq:w_shrink}
    \W = \frac{1}{\tau^{\pilot}}\R(\alpha)^{-1} \Y^{\pilot}\P.
\end{align}

Several methods have been proposed in the literature for computing the shrinkage coefficient to enhance the estimation accuracy of the covariance matrix~\cite{Che11}. A simple and well-known technique is based on the oracle estimator, where the shrinkage coefficient is obtained by minimizing the difference between the regularized sample covariance matrix of the received pilot signal and the actual covariance matrix, i.e.,
\begin{align}
    & \underset{\alpha}{\argmin} \left\| \R(\alpha) - (\H \Omegab_{\rho} \H^{\herm} + \Psib)\right\|_{\fro}^2, \nonumber \\
    & = \frac{1}{\tr(\S \S^{\herm})} \Re \left[ \tr\bigg( \bigg(\H \Omegab_{\rho} \H^{\herm} + \Psib - \Q \bigg) \S \bigg) \right]. \label{eq:or_sc}
\end{align}
However, this approach is obviously impractical as it requires the knowledge of $\H$ and $\Psib$. Therefore, we propose an alternative approach to optimize the shrinkage coefficient based on the received data signal in Section~\ref{sec:reg1}. Additionally, the criterion in \eqref{eq:or_sc} based on the received pilot signal may not directly optimize metrics such as the SER or SINR~\cite{Kuz16}. Hence, we propose to optimize the shrinkage coefficient by iteratively minimizing the sample MSE of the hard-decision symbols in Section~\ref{sec:reg2}.
 
\subsection{Regularization Based on the Received Data Signal} \label{sec:reg1}
The oracle estimator in \eqref{eq:or_sc} is infeasible due to the required knowledge of $\H$ and $\Psib$. However, since the number of data symbols is generally larger than the number of pilot symbols, i.e., \( \tau^{\data} \ge \tau^{\pilot} \), we can utilize the sample covariance matrix of the received data signal $\frac{1}{\tau^{\data}} \Y^{\data} (\Y^{\data})^{\herm}$, which also converges to $\H \Omegab_{\rho} \H^{\herm} + \Psib$ as \( \tau^{\data} \rightarrow \infty \). Consequently, the shrinkage coefficient can be obtained by minimizing the difference between the regularized sample covariance matrix of the received pilot signal and the sample covariance matrix of the received data signal, i.e.,
\begin{align}
    & \underset{\alpha}{\argmin} \left\|\R(\alpha) - \frac{1}{\tau^{\data}} \Y^{\data} (\Y^{\data})^{\herm}\right\|_{\fro}^2 \nonumber \\
    & = \frac{1}{\tr(\S \S^{\herm})} \Re \left[ \tr\left( \left( \frac{1}{\tau^{\data}} \Y^{\data} (\Y^{\data})^{\herm} - \Q \right) \S \right) \right].
\end{align}

Indeed, the sample covariance matrix of the received data signal can also be utilized directly to enhance the estimation accuracy of the covariance matrix by employing alternative objective functions, as detailed in~\cite{Oli21}. However, the estimation accuracy may be poor in interference-limited scenarios. Moreover, improving the estimation accuracy of the covariance matrix may not directly translate into improvements in metrics such as SER and SINR. Therefore, we propose an iterative optimization of the shrinkage coefficient based on the sample MSE of the hard-decision symbols, which is more closely related to the system's performance in terms of SER and~SINR.

\subsection{Iterative Regularization Based on the Hard-Decision Symbols} \label{sec:reg2}

In this method, the shrinkage coefficient in \eqref{eq:cov_shrink} is iteratively updated using the hard-decision symbols. Specifically, at each iteration and for a fixed shrinkage coefficient \( \alpha \), the soft-estimated symbols of UE~$k$ are computed as in \eqref{eq:d_k_est} using the direct-estimate combiner of UE~$k$, which corresponds to the $k$th column of  \( \W \) in \eqref{eq:w_shrink}. The hard-decision symbols are then obtained by mapping the soft-estimated symbols to the set of transmitted data symbols using the minimum-distance criterion, which offers lower complexity than the optimal maximum-likelihood detector and is expressed as
\begin{align}\label{eq:d_bar}
    \bar{\d}_k \triangleq \argmin_{\d_k \in \setD_k^{\tau^{\data} \times 1}} \| \hat{\d}_k - \d_k \|_2^2,
\end{align}
where \( \setD_k \) is the set of all possible transmitted symbols of UE~$k$. Then, using $\bar{\d}_k$, the shrinkage coefficient $\alpha$ is optimized by minimizing the sample MSE between the soft-estimated symbols in \eqref{eq:d_k_est} and the hard-decision symbols in \eqref{eq:d_bar}. To this end, we express the average sample MSE across all the UEs~as
\begin{align}
    \epsilon(\alpha) & \triangleq \! \frac{1}{K \tau^{\data}} \sum_{k \in \setK} \big\|(\Y^{\data})^{\herm}  \w_k - \bar{\d}_k\big\|_{2}^2 \nonumber \\
    & = \! \frac{1}{K \tau^{\data}} \tr\bigg( \! \frac{1}{({\tau^{\pilot}})^2}\P^{\herm} (\Y^{\pilot})^{\herm} \R(\alpha)^{-1} \Y^{\data} (\Y^{\data})^{\herm} \R(\alpha)^{-1} \Y^{\pilot}\P \nonumber \\
    & \phantom{=} \ \! - \frac{2}{\tau^{\pilot}}\Re\big[\P^{\herm} (\Y^{\pilot})^{\herm} \R(\alpha)^{-1} \Y^{\data} \bar{\D}^{\herm}\big] + \bar{\D} \bar{\D}^{\herm}\bigg), \label{eq:er_sam_MSE}
\end{align}
with $\bar \D \triangleq [\bar \d_1, \ldots, \bar \d_K]^{\tran} \in \Compl^{ \tau^{\data} \times K}$. In \eqref{eq:er_sam_MSE}, minimizing $\epsilon(\alpha)$ with respect to $\alpha$ can be interpreted as minimizing the cross-validation loss with respect to the ridge parameter in ridge regression~\cite{Wil21}. Since $\epsilon(\alpha)$ is not convex with respect to $\alpha$, we aim to achieve a locally optimal $\alpha$ using a gradient-based method. For this purpose, we write the derivative of $\epsilon(\alpha)$ with respect to $\alpha$ as
\begin{align}\label{eq:alpha_gupd}
    \frac{\partial}{\partial \alpha} \epsilon(\alpha) &= - \frac{2}{K \tau^{\data}}\Re \bigg[\tr\bigg( \P^{\herm} (\Y^{\pilot})^{\herm} \R(\alpha)^{-1} \S  \R(\alpha)^{-1} \Y^{\data} \nonumber \\
     & \phantom{=} \ \times \bigg( \frac{1}{(\tau^{\pilot})^2}(\Y^{\data})^{\herm}  \R(\alpha)^{-1}  \Y^{\pilot}\P  + \frac{1}{\tau^{\pilot}}\bar \D^{\herm} \bigg) \bigg)\bigg].
\end{align}
At each iteration $i$, $\alpha$ can be updated in the direction of the negative derivative as 
\begin{align}\label{eq:alpha_ud}
    \alpha^{(i)} =  \alpha^{(i-1)} - \beta \frac{\partial}{\partial \alpha} \epsilon (\alpha) \bigg|_{\alpha=\alpha^{(i-1)}},
\end{align}
where $\beta > 0$ is the step size chosen to promote faster convergence. The iterative update of the shrinkage coefficient $\alpha$ based on the hard-decision symbols is outlined in Algorithm~\ref{alg:alpha_ud}, where an example of predefined termination criterion is that the value of $\alpha$ does not change significantly between consecutive iterations.

\begin{figure}[t!]
\begin{algorithm}[H]
\small
\begin{spacing}{1.1}
\textbf{Data:} $\P \in \Compl^{\tau^{\pilot} \times K}$, $\Y^{\pilot}  \in \Compl^{BM \times \tau^{\pilot}}$, and $\Y^{\data}  \in \Compl^{BM \times \tau^{\data}}$.\\
\textbf{Initialization:} $\alpha^{(0)} = 0$, $i = 0$, and $\beta$.\\
\textbf{Until} a predefined termination criterion is satisfied, \textbf{do:}
\begin{itemize}
    \item[\texttt{1)}] $i \leftarrow i + 1$.
    \item[\texttt{2)}] Compute $\W$ as in \eqref{eq:w_shrink} with $\alpha = \alpha^{(i-1)}$.
    \item[\texttt{3)}] Compute $\hat{\d}_k$ as in \eqref{eq:d_k_est} for all the UEs.
    \item[\texttt{4)}] Compute $\bar{\d}_k$ as in \eqref{eq:d_bar} for all the UEs.
    \item[\texttt{5)}] Compute $\frac{\partial \epsilon(\alpha)}{\partial \alpha}$ as in \eqref{eq:alpha_gupd} and update $\alpha^{(i)}$ as in \eqref{eq:alpha_ud}.
\end{itemize}
\textbf{End}
\end{spacing}
\caption{Iterative update of the shrinkage coefficient}
\label{alg:alpha_ud}
\end{algorithm}
\vspace{-6mm}
\end{figure}

We point out that \( \mathbf{R}(\alpha) \) in \eqref{eq:alpha_gupd} does not need to be inverted at each iteration. In fact, since \( \mathbf{R}(\alpha) \) retains the same eigenvectors as \( \mathbf{Q} \) for any \( \alpha \), one can compute the eigenvalue decomposition of \( \mathbf{Q} \) at the beginning of the algorithm and update only the eigenvalues at each iteration, simplifying the inversion of \( \mathbf{R}(\alpha) \) to that of a diagonal matrix.
Additionally, a stochastic gradient-based method can be employed to compute the derivative in \eqref{eq:alpha_gupd} based on a subset of received data symbols, further reducing the computational complexity.

\section{Numerical Results}
We consider a distributed MIMO system with \( B = 2 \) APs, each equipped with \( M = 4 \) antennas and placed $100$~m apart, serving \( K = 6 \) single-antenna UEs. Assuming uncorrelated Rayleigh fading, each channel is generated as \( \mathrm{vec}(\H_{b,k}) \sim \setC \setN(0, \zeta_{b,k} \I_{MN}) \), with \( \zeta_{b,k} \triangleq -30.5 -36.7 \log_{10}(r_{b,k}) \) [dB], where \( r_{b,k} \) denotes the distance between AP~\( b \) and UE~\( k \)~\cite{Atz21}. The AWGN power at the APs is fixed to \(-95\)~dBm. The proposed regularization methods in Section~\ref{sec:reg1} and Section~\ref{sec:reg2}  (Algorithm~\ref{alg:alpha_ud}) are referred to as \textit{Reg.~data} and \textit{Reg.~data~iter.}, respectively. These methods are compared with: \textit{i)} no regularization (\( \alpha = 0 \)), referred to as \textit{No~reg.}; \textit{ii)} regularization optimized through exhaustive line search to minimize the MSE, referred to as \textit{Reg.~exh.}; and \textit{iii)} the ideal case with perfect CSI, referred to as \textit{Perfect~CSI}.

\begin{figure}[t!]
\begin{center}
\begin{tikzpicture}

\begin{axis}[
	width=8cm,
	height=6cm,
	xmin=0, xmax=22,
	ymin=1e-4, ymax=1,
    xlabel={$\rho_k$ [dBm]},
    ylabel={SER },
	xtick={2,6,10,14,18,22},
    xlabel near ticks,
	ylabel near ticks,
    x label style={font=\footnotesize},
	y label style={font=\footnotesize},
    ticklabel style={font=\footnotesize},
    legend pos=south east,
    legend cell align=left,
    legend style={at={(0.01,0.01)}, anchor=south west},
	legend style={font=\scriptsize, inner sep=1pt, fill opacity=0.75, draw opacity=1, text opacity=1},
	grid=both,
 ymode=log,
    log basis y={10}
]

\addplot[dashed, line width=1pt, magenta, mark=triangle, mark options={solid}]
table[x expr={30+\thisrow{pwrdB}}, y=DE,  , col sep=comma] 
{Figs/fig1data.txt};
\addlegendentry{{No reg.}};

\addplot[line width=1pt, red, mark=o, mark options={solid}]
table[x expr={30+\thisrow{pwrdB}}, y=dataCov, col sep=comma] 
{Figs/fig1data.txt};
\addlegendentry{{Reg. data}};

\addplot[line width=1pt,  blue, mark=x, mark options={solid}]
table[x expr={30+\thisrow{pwrdB}}, y=DEgrd,  , col sep=comma] 
{Figs/fig1data.txt};
\addlegendentry{{Reg. data iter.}};

\addplot[dashed, line width=1pt, green, mark=diamond, mark options={solid}]
table[x expr={30+\thisrow{pwrdB}}, y=DEsearch,  , col sep=comma] 
{Figs/fig1data.txt};
\addlegendentry{{Reg. exh.}};

\addplot[dotted, line width=1pt, black, mark=+, mark options={solid}]
table[x expr={30+\thisrow{pwrdB}}, y=Idl, col sep=comma] 
{Figs/fig1data.txt};
\addlegendentry{{Perfect CSI}};

\end{axis}

\end{tikzpicture}
\vspace{-3mm}
\caption{SER of QPSK versus UE transmit power for \( K = 6 \), \( \tau^{\pilot} = 8 \), and \( \tau^{\data} = 1000 \), without interference.}
\label{fig:noInt}
\end{center}
\vspace{-4mm}
\end{figure}

In Figure~\ref{fig:noInt}, the SER for QPSK is plotted against the UE transmit power, without any external interference source. Both \textit{Reg.~data} and \textit{Reg.~data~iter.} outperform \textit{No~reg.}, with \textit{Reg.~data~iter.} achieving a gain of $3$--$4$~dB compared to \textit{No~reg.} and performing closely to \textit{Reg.~exh.}. However, at low UE transmit powers, the performance of \textit{Reg.~data~iter.} is inferior to \textit{Reg.~data} due to errors in the hard-decision symbols. As the UE transmit power increases, the SER performance of \textit{Reg.~data~iter.} surpasses that of \textit{Reg.~data}, demonstrating considerable improvements for a fixed UE transmit power.

In Figure~\ref{fig:Int}, the SER for QPSK is plotted against the UE transmit power, with an external interference source. The power of the interference source is set to be $5$~dB lower than the UE transmit power, and it transmits random symbols during both the pilot and data transmission phases. The behavior of \textit{No~reg.}, \textit{Reg.~data~iter.}, and \textit{Reg.~exh.} is similar to the results shown in Figure~\ref{fig:noInt}. However, regularization using \textit{Reg.~data}, which aligns the sample covariance matrices obtained from the received pilot and data signals, yields a suboptimal shrinkage coefficient. This is because the signals from random interference sources during the pilot phase exhibit higher correlation due to the limited sample size compared with the data transmission phase. This leads to a degradation in the SER performance, even compared with \textit{No~reg.}

\begin{figure}[t!]
\begin{center}
\begin{tikzpicture}

\begin{axis}[
	width=8cm,
	height=6cm,
	xmin=0, xmax=22,
	ymin=1e-4, ymax=1,
    xlabel={$\rho_k$ [dBm]},
    ylabel={SER },
	xtick={2,6,10,14,18,22},
    xlabel near ticks,
	ylabel near ticks,
    x label style={font=\footnotesize},
	y label style={font=\footnotesize},
    ticklabel style={font=\footnotesize},
    legend pos=south east,
    legend cell align=left,
    legend style={at={(0.01,0.01)}, anchor=south west},
	legend style={font=\scriptsize, inner sep=1pt, fill opacity=0.75, draw opacity=1, text opacity=1},
	grid=both,
 ymode=log,
    log basis y={10}
]

\addplot[dashed, line width=1pt, magenta, mark=triangle, mark options={solid}]
table[x expr={30+\thisrow{pwrdB}}, y=DE,  , col sep=comma] 
{Figs/fig2data.txt};
\addlegendentry{{No reg.}};

\addplot[line width=1pt, red, mark=o, mark options={solid}]
table[x expr={30+\thisrow{pwrdB}}, y=dataCov, col sep=comma] 
{Figs/fig2data.txt};
\addlegendentry{{Reg. data}};

\addplot[line width=1pt,  blue, mark=x, mark options={solid}]
table[x expr={30+\thisrow{pwrdB}}, y=DEgrd,  , col sep=comma] 
{Figs/fig2data.txt};
\addlegendentry{{Reg. data iter.}};

\addplot[dashed, line width=1pt, green, mark=diamond, mark options={solid}]
table[x expr={30+\thisrow{pwrdB}}, y=DEsearch,  , col sep=comma] 
{Figs/fig2data.txt};
\addlegendentry{{Reg. exh.}};

\addplot[dotted, line width=1pt, black, mark=+, mark options={solid}]
table[x expr={30+\thisrow{pwrdB}}, y=Idl, col sep=comma] 
{Figs/fig2data.txt};
\addlegendentry{{Perfect CSI}};

\end{axis}

\end{tikzpicture}
\vspace{-3mm}
\caption{SER of QPSK versus UE transmit power for \( K = 6 \), \( \tau^{\pilot} = 8 \), and \( \tau^{\data} = 1000 \), with interference.}
\label{fig:Int}
\end{center}
\vspace{-4mm}
\end{figure}

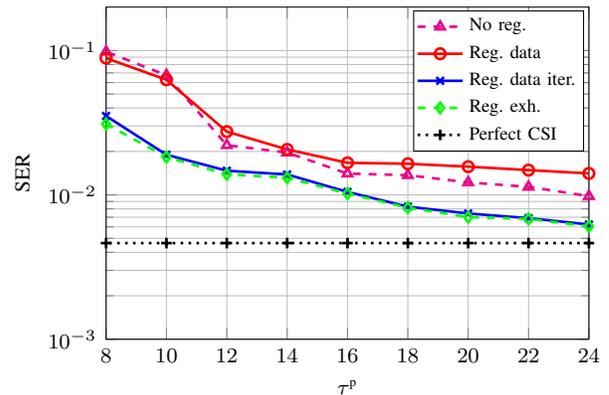
\begin{figure}[t!]
\begin{center}
\begin{tikzpicture}

\begin{axis}[
	width=8cm,
	height=6cm,
	xmin=8, xmax=24,
	ymin=1e-3, ymax=0.2,
    xlabel={$\tau^{\pilot}$},
    ylabel={SER },
    xlabel near ticks,
	ylabel near ticks,
    x label style={font=\footnotesize},
	y label style={font=\footnotesize},
    ticklabel style={font=\footnotesize},
    legend pos=south east,
    legend cell align=left,
    legend style={at={(0.99,0.99)}, anchor=north east},
	legend style={font=\scriptsize, inner sep=1pt, fill opacity=0.75, draw opacity=1, text opacity=1},
	grid=both,
 ymode=log,
    log basis y={10}
]

\addplot[dashed, line width=1pt, magenta, mark=triangle, mark options={solid}]
table[x=pltLen, y=DE,  , col sep=comma] 
{Figs/fig3data.txt};
\addlegendentry{{No reg.}};

\addplot[line width=1pt, red, mark=o, mark options={solid}]
table[x=pltLen, y=dataCov, col sep=comma] 
{Figs/fig3data.txt};
\addlegendentry{{Reg. data}};

\addplot[line width=1pt,  blue, mark=x, mark options={solid}]
table[x=pltLen, y=DEgrd,  , col sep=comma] 
{Figs/fig3data.txt};
\addlegendentry{{Reg. data iter.}};

\addplot[dashed, line width=1pt, green, mark=diamond, mark options={solid}]
table[x=pltLen, y=DEsearch,  , col sep=comma] 
{Figs/fig3data.txt};
\addlegendentry{{Reg. exh.}};

\addplot[dotted, line width=1pt, black, mark=+, mark options={solid}]
table[x=pltLen, y=Idl, col sep=comma] 
{Figs/fig3data.txt};
\addlegendentry{{Perfect CSI}};

\end{axis}

\end{tikzpicture}
\vspace{-3mm}
\caption{SER of QPSK versus pilot length for \( K = 6 \), \( \rho_k = 15 \)~dBm, and \( \tau^{\data} = 1000 \),  with interference.}
\label{fig:SERvsPlt}
\end{center}
\vspace{-6mm}
\end{figure}

In Figure~\ref{fig:SERvsPlt}, the SER for QPSK is plotted against the pilot length with an external interference source. The interference characteristics are similar to those used in Figure~\ref{fig:Int}. As the pilot length increases, the performance of \textit{No~reg.}, \textit{Reg.~data~iter.}, \textit{Reg.~exh.}, and \textit{Reg.~data} improves, approaching the SER of \textit{Perfect~CSI}. However, a significant SER gap remains between \textit{Reg.~data~iter.} and both \textit{No~reg.} and \textit{Reg.~data}. Furthermore, the performance of \textit{Reg.~data~iter.} is close to that of \textit{Reg.~exh.} for any pilot length.

\vspace{-1mm}
\section{Conclusions}
Considering the uplink of a distributed MIMO system, we investigated the data-aided regularization of the direct-estimate combiner to address the inaccurate estimation of the covariance matrix used in its computation due to the limited number of pilot symbols. Specifically, we applied a shrinkage-based regularization method and initially optimized the shrinkage coefficient by minimizing the difference between the covariance matrices obtained from the received pilot and data signals. Recognizing the limitations of this approach in interference-limited scenarios, we proposed an iterative method to optimize the shrinkage coefficient by minimizing the sample MSE of the hard-decision symbols. Numerical results demonstrated that the proposed methods significantly improve the SER, particularly under interference-limited conditions and with a small number of pilot symbols.

\bibliographystyle{IEEEtran}
\bibliography{IEEEabbr,refs}
\end{document}